# Curie Temperature of Emerging Two-Dimensional Magnetic Structures

Xiaobo Lu[1], Ruixiang Fei[1], and Li Yang[1,2]

[1]Department of Physics, Washington University in St. Louis, St. Louis, MO 63130, USA.
[2]Institute of Materials Science and Engineering, Washington University in St. Louis, St. Louis, MO 63130, USA.

Abstract

Recent realizations of intrinsic, long-range magnetic orders in two-dimensional (2D) van der Waals materials have ignited tremendous research interests. In this work, we employ the XXZ Heisenberg model and Monte Carlo simulations to study a fundamental property of these emerging 2D magnetic materials, the Curie temperature ($T_c$). By including both onsite and neighbor couplings extracted from first-principles simulations, we have calculated $T$c of monolayer chromium trihalides and $Cr_2Ge_2Te_6$, which are of broad interests currently, and the simulation results agree with available measurements. We also clarify the roles played by anisotropic and isotropic interactions in deciding $T_c$ of magnetic orders. Particularly, we find a universal, linear dependence between $T_c$ and magnetic interactions within the parameter space of realistic materials. With this linear dependence, we can predict $T_c$ of general 2D lattice structures, omitting the Monte Carlo simulations. Compared with the widely used Ising model, mean-field theory, and spin-wave theory, this work provides a convenient and quantitative estimation of $T_c$, giving hope to speeding up the search for novel 2D materials with higher Curie temperatures.

**I. Introduction**

Long-range magnetic orders had been believed to hardly survive in two-dimensional (2D) systems due to the enhanced thermal fluctuations that make spontaneously symmetry-breaking orders unsustainable [1,2]. In 1960s, Mermin-Wagner (MW) theorem showed that such symmetry-breaking order states are only ruled out if considering continuous rotational symmetries and short-range interactions, indicating the existence of 2D magnetism with anisotropic interactions [3–5]. The recent discoveries of 2D magnetic crystals confirmed that magnetic anisotropy plays a crucial role in realizing 2D ferromagnetism [5–8]. With the help of magnetic anisotropy, the long-range ferromagnetic (FM) order could be established in 2D structures by opening a magnon gap to resist the thermal agitation [7,9–11]. In early 2017, the ferromagnetic (FM) order in pristine 2D crystals was observed in both monolayer $CrI_3$ and $Cr_2Ge_2Te_6$ (CGT), and great enthusiasm has been aroused for searching and exploring 2D magnetism. [6,7] More recently, many other 2D magnetic materials, *e.g.*, $FePS_3$ [12,13], $Fe_3GeTe_2$ [14,15], $VSe_2$ [16] $MnSe_2$ [17], and $MnBi_2Te_4$ [18,19], have been realized by either exfoliation from bulk structures or growth with molecular beam epitaxy (MBE). Interesting magnetic properties, such as room-temperature intrinsic ferromagnetism [16], magnetic topological insulators [18,19], and electric-field/doping tunable magnetism [20–22], etc. have been observed in these newborn 2D magnetic materials.

The most important character of ferromagnetism is the phase-transition temperature, *i.e.*, the Curie temperature ($T_c$), which not only decides applications but also reflects the intrinsic magnetic mechanism. To date, numerous theoretical calculations have focused on this important magnetic property [23]. Take the intensively studied monolayer $CrI_3$ as an example. The experimentally measured value is around 45K [6]. The Ising model predicted an overrated $T_c$ over 80 K [24,25] because of the overestimation of anisotropic interactions. The spin-wave theory gave a better estimation of 33 K [25]. However, the harmonic and mean-field approximations in the spin-wave Hamiltonian and the corresponding magnetization introduce extra error bars. A few Monte Carlo (MC) simulations predicted $T_c$ in a range between 50K and 96K, depending on their parameters and Hamiltonians [26,27]. Importantly, recent studies [28–30] not only show the reliability of MC simulations comparing with the random phase approximation, and further reveal direct relations between $T_c$ and corresponding magnetic interaction strengths. Therefore, along this direction, more works are necessary for theoretically calculating reliable $T_c$ of those 2D magnetic structures and understanding anisotropic magnetic interactions and their impacts on 2D magnetism.

In this work, we start from the XXZ Heisenberg model with the magnetic interaction parameters extracted from first-principles simulations. By using the MC simulation, we obtain the $T_c$ of the widely studied monolayer chromium trihalides and CGT. Our result of monolayer $CrI_3$ is around 42K that is in an excellent agreement with the measurement (45K). More interestingly, we find that there is a universal, linear dependence between $T_c$ and magnetic interaction coefficients within a wide range of parameter space. As a result, we can conveniently predict $T_c$ of 2D Heisenberg-type magnetic materials based on the magnetic interactions without time-consuming MC simulations. Using the MC-simulated magnetic phase diagram, we explain the origin of such a linear relation and show the crucial role of anisotropic magnetic interactions in creating and keeping 2D magnetism. This advance can be useful for quantitatively understanding the origin of 2D magnetism and speeding up the discovery of novel 2D magnetic materials.

The article is organized in the following order. In section II, we present the atomic structures of our studied 2D magnetic materials and first-principles simulation setups. In Section III, we introduce the XXZ Heisenberg model and MC simulations. In section IV, the MC simulation results based on the XXZ Heisenberg model are presented and compared with measurements. In Section V, we reveal the linear dependence of the $T_c$ on magnetic interactions. In section VI, we discuss the phase diagram of 2D magnetism according to the anisotropic onsite and exchange interactions to understand the linear dependence and the roles played by anisotropic magnetic interactions. In Section VII, we summarize the results.

## II. Atomic Structure and First-principles Simulation Setups

*DFT calculations:* The DFT calculations are performed within the generalized gradient approximation (GGA) using the Perdew-Burke-Ernzerhof (PBE) functional implemented in Vienna Ab initio Simulation Package (VASP). [31,32] A plane-wave basis set with a kinetic energy cutoff of 450eV, and a 5x5x1 k-point sampling grid is adopted for a 2x2x1 supercell to mimic different magnetic configurations for extracting magnetic interactions. The vacuum distance is set to be 20 Å between adjacent layers to avoid spurious interactions. The van der Waals (vdW) interaction is included by the DFT-D2 method, [33] and spin-orbit coupling (SOC) is always considered. We choose the Hubbard U = 2.7eV and Hund J = 0.7eV parameters for $Cr^{3+}$ ions, which has been widely used in previous works. [24,25,34] The geometric structure is relaxed until the force converged within 0.01 eV/Å.

*Atomic Structure:* Monolayer chromium trihalides and CGT share a similar lattice structure. Take monolayer $CrI_3$ as an example. As shown in Figure 1 (a), $Cr^{3+}$ cations are arranged in honeycomb lattices while carrying the localized magnetic moments (3μ$_B$/Cr$^{3+}$), and they are coordinated to six nearest-neighbor I$^-$ anions, forming edge-sharing octahedral. By maintaining the $C_3$ rotational symmetry, our fully relaxed in-plane lattice constant is 6.91 Å, which well agrees with previous published results. [24,27]

**III. Heisenberg model and MC simulation setup**

The Heisenberg model is a widely employed approach to study 2D magnetism since the early works by Mermin and Wagner [1–4]. The Heisenberg formulism provides enriched stages for various physics phenomenon in 2D magnetism, such as quantum critical behavior and none-trivial phase transitions, *i.e.*, the Berezinskii-Kosterlitz-Thouless (BKT) transition in 2D Heisenberg model with easy-plane anisotropy. [3,4,35,36] Previous works showed that 2D magnetic materials, such as monolayer CrI$_3$, exhibit an easy axis along the out-of-plane direction, and the magnetic energy is approximately isotropic along in-plane directions, making it reasonable to further mutate the Heisenberg model into a quantum anisotropic Heisenberg model, also called the XXZ model. [6,25,37,38]

In this work, we construct the XXZ Hamiltonian by including both onsite and coupling anisotropic magnetic interactions as following:

$$\mathcal{H} = \sum_i A(S_i^z)^2 + \sum_{<i,j>} \frac{1}{2}\left(\lambda_1 S_i^z S_j^z + J_1 \vec{S}_i \cdot \vec{S}_j\right) + \sum_{\ll i,j \gg} \frac{1}{2}\left(\lambda_2 S_i^z S_j^z + J_2 \vec{S}_i \cdot \vec{S}_j\right) \quad (1)$$

As indicated in Figure 1 (b), the interactions among those highly localized magnetic moments can be reasonably described by neighbor couplings which include both isotropic exchange interaction *J* and anisotropic exchange coupling λ. The subscript 1 means the nearest neighbor (NN) coupling, and the subscript 2 means the next NN (NNN) coupling. The coefficient *A* describes the easy-axis, single-ion anisotropy. In this work, we neglect the 3rd NN and farther couplings because they are an order of magnitude smaller. [27,38]

To obtain the phase transition and $T_c$, we perform MC simulations based on the Metropolis algorithm on 2D hexagonal lattices with a size of 40x40 unit cells, which contain 3200 magnetic moments. The periodic boundary condition is implemented. A MC step consists of an attempt to assign a new random direction in 3D space to one of random magnetic moments in lattices. All magnetic moments point along the out-of-plane direction at the initial state to mimic experimental setups, in which the low-temperature ground state is obtained under external assisting magnetic field. [6,7,39,40] We run for $4 \times 10^8$ MCs ($2.5 \times 10^5$ steps per site average) to ensure that the thermal equilibrium is achieved. For each temperature point, there are 20 independent runs to reduce the statistical fluctuation.

The magnetization is defined as

$$<m_\alpha> = \frac{1}{N}\sum_{i=1}^{N} <S_i^\alpha> \quad (2),$$

where $N$ represents the total magnetic moments in the simulated system, and $<S_i^\alpha>$ indicates the time average of corresponding magnetic components after the simulation achieves thermal equilibrium. Finally, $T_c$ can be estimated by fitting the typical phase transition formula:

$$<m_z> = \begin{cases} \mu(Tc-T)^\delta & , Tc > T \\ 0 & , Tc < T \end{cases} \quad (3)$$

**IV. Magnetic phase transition and $T_c$**

First, we must obtain the coefficients of magnetic interactions in the Hamiltonian of Eq (1) by calculating total energies of different magnetic configurations. Here we consider the FM and Néel antiferromagnetic (AFM) configurations. The corresponding energy expressions of a unit cell are

$$E_{FM/AFM}^{out} = E_0 + (2A \pm 3\lambda_1 \pm 3J_1 + 6\lambda_2 + 6J_2)S^2 \quad (4),$$

$$E_{FM/AFM}^{in} = E_0 + (\pm 3J_1 + 6J_2)S^2 \quad (5),$$

in which two magnetic orientations (in-plane and out-of-plane ones) are calculated in order to specify anisotropic couplings. Moreover, we can flip the magnetic moment of a $Cr^{3+}$ cation in a 2x2x1 supercell to obtain more energy configurations (normalized to one unit cell).

$$E_{flip}^{out} = E_0 + \left(2A + \frac{3}{2}\lambda_1 + \frac{3}{2}J_1 + 3\lambda_2 + 3J_2\right)S^2 \quad (6)$$

$$E_{flip}^{in} = E_0 + \left(\frac{3}{2}J_1 + 3J_2\right)S^2 \quad (7)$$

As a result, we accumulate six equations that can solve the five magnetic interaction coefficients in Eq (1) and reference energy ($E_0$).

The magnetic interaction coefficients, which are extracted from DFT calculations, of monolayer chromium trihalides are summarized in Table I. As expected, the isotropic NN coupling ($J_1$) is the strongest and has a negative sign, resulting in the FM ground state. Both the signs of the anisotropic NN coupling ($\lambda_1$) and onsite anisotropic term ($A$) are negative, indicating the easy axis of magnetization is along the z (out-of-plane) direction. Importantly, anisotropic interactions are significantly weaker than isotropic ones. For example, the isotropic NN coupling $J_1$ is about -2.12 meV for monolayer $CrI_3$, and the anisotropic NN coupling $\lambda_1$ is about -0.085 meV. These results are close to previous published results [25]. Finally, as discussed in previous works, magnetic interactions of these monolayer structures are mainly from halogen-atom SOC mediated super-exchange interactions between $Cr^{3+}$ cations [25,41]. Therefore, we observe that magnetic interactions are reduced from $CrI_3$, $CrBr_3$, to $CrCl_3$, along the trend of lighter halogen atoms.

With these magnetic coupling coefficients, we can obtain the dependence of magnetization according to temperature by MC simulations in Figure 2. Magnetic phase transitions are observed, and $T_c$ is 42.2 K, 23.1 K, and 12.1 K for monolayer $CrI_3$, $CrBr_3$, and $CrCl_3$, respectively. To date, the measured Curie temperatures of monolayer $CrI_3$ and $CrBr_3$ are around 45K and 27K, respectively [6,42] Compared with previous results from the Ising model, mean-field theory, and spin-wave theory, our calculation agrees better with measurements. For $CrCl_3$, the available measurement of the 2L structure is about 17 K which is a slightly higher than we monolayer result $T_c$ (12K). This enhancement of $T_c$ of bilayer may be from interlayer couplings [42,43] We have to address that these results are sensitive to the choices of U and J. As shown in in previous works, [28–30] different values of these two parameters will change the calculated $T_c$ because the magnetic interactions are changed accordingly.

It must be pointed out that, although the NNN coupling $J_2$ is significantly smaller than $J_1$, its contribution to $T_c$ is not small due to the larger coordinate number of the next NN in hexagonal lattices ($N_2 = 6$). For chromium trihalides, we find that including the NNN coupling raises $T_c$ by roughly 20%-30%. It is also worth to point out that the anisotropy of both onsite energy and exchange interaction in $CrCl_3$ is quite weak. In such weakly anisotropic cases, the help of a small external field may be needed to observe the FM phase transition. This is similar to what has been discussed in the CGT experiment [7].

## V. Relationships between $T_c$ and magnetic interaction

Although MC simulations provide a good solution to the XXZ Hamiltonian, it is time consuming and complicated for spreading out to specific materials that uses combinations of parameters. Since the magnetic energy and $T_c$ are fundamentally decided by magnetic interactions, it will be interesting to explore if there is a direct relation between $T_c$ and those interaction coefficients.

For this purpose, we scan a much larger parameter space by MC simulations based on the XXZ Heisenberg model (Eq. 1). However, we have five parameters, *i.e.*, $A$, $J_1$, $J_2$, $\lambda_1$ and $\lambda_2$. It is impossible to cover such a five-dimensional parameter space, and we must simplify the scanning space. For most magnets, the NN isotropic coupling $J_1$ is usually the dominant factor for the amplitude of $T_c$. Therefore, we carry out extensive MC simulations by varying $J_1$ and one of other parameters from $A$, $\lambda_1$, $\lambda_2$ and $J_2$ while fixing the rests (e.g., at values of monolayer $CrI_3$). The results are showing in Figure 3. Surprisingly, we find that $T_c$ for each $J_1$ vs. $\alpha$ ($\alpha$=A, $\lambda_1$, $\lambda_2$, $J_2$) is roughly sitting in a flat plane. It indicates that $T_c$ of such an XXZ Heisenberg model is linearly dependent with these coupling strengths. This qualitatively agrees with the expressions in Ref. [28,29], which are based on the Ising limit value with NN couplings included.

$T_c$ can be expressed as

$$k_B T_c = (\alpha_0) * AS^2 + (\alpha_1^z * n_1) * \lambda_1 S^2 + (\alpha_1 * n_1) * J_1 S^2$$
$$+ (\alpha_2^z * n_2) * \lambda_2 S^2 + (\alpha_2 * n_2) * J_2 S^2 \qquad (8).$$

The NN and NNN coordinate number, $n_i$, and fitted values, $\alpha_i$, are listed in Table II. With this linear dependence, we can predict $T_c$ of 2D magnets without the MC simulation. As shown in Table I, this model predicted $T_c$ is very close to MC results with an error bar of less than 1 K for monolayer chromium trihalides. We have tried this model to another family of 2D magnets, CGT, which has the similar hexagonal structures as chromium trihalides. Using the published magnetic interaction coefficients [7], our estimated $T_c$ of monolayer CGT is 34.5 K, which is kinda close to the available measured value (around 30 K under a 0.075T external field) of bilayer CGT. [7] It has to be pointed out that, according to the MW theorem, this expression is no longer valid when all the anisotropy parameters approach zero. Fortunately, such an invalid region is small owing to the logarithmically asymptotical behavior of $T_c$, which will be discussed in section VI.

Beyond hexagonal lattices, we also imitate the same processing of MC simulations on 2D square lattices by the XXZ Hamiltonian. As showing in Figure 4, $T_c$ for each $J_1$ vs. α (α=A, $\lambda_1$, $\lambda_2$, $J_2$) is roughly sitting in a flat plane. Therefore, the similar linear dependence is concluded, and the extracted values of parameters of square lattices are listed in Table II. Interestingly, the coefficients of the linear model for both square and hexagonal lattices are similar as shown in Table II. Therefore, this linearity is robust for different lattices, and the $T_c$ expression (Eq. 8) could be universal for anisotropic 2D XXZ Heisenberg systems.

In fact, a linear dependence of $T_c$ according to the magnetic coupling coefficients has been proposed in other-level models. [44] For instance, it can derived from the Hamiltonian in Eq. (1) that the widely used mean-field approximation (MFA) gives the linear relation [44]:

$$T_c = \frac{2S(S+1)}{3k_B}\left(-A - \frac{n_1}{2}*J_1 - \frac{n_1}{2}*\lambda_1 - \frac{n_2}{2}*J_2 - \frac{n_2}{2}*\lambda_2\right) \qquad (9)$$

However, there are several obvious deficiencies in that MFA result. Firstly, the MFA cannot distinguish isotropic and anisotropic couplings due to that all operators are replaced by their thermodynamic mean values along the out-of-plane direction. As a result, the slops of the linear dependence are same for both isotropic $J$ and anisotropic $\lambda$. This overestimated anisotropy is a main reason for the much larger $T_c$ provided by the MFA. In other words, MFA shall be appropriate for very strongly anisotropic magnetic systems

Second, $T_c$ is proportional to S(S+1) in the MFA but our model shows that $T_c$ is proportional to $S^2$. This discrepancy origins from the different quantum and classic treatments of the spin operator. To verify this point, we change the magnetic moment in MC simulations. Meanwhile, we tune the coupling constants simultaneously to keep the value of the product of $\alpha S_i^{(z)} \cdot S_j^{(z)}$ ($\alpha = A, \lambda, J$). In Figure 5, the simulation shows that $T_c$ does not change for different magnetic moments. For instance, the curve starts from $3\mu_B$ is simulating monolayer CrI$_3$ which has a magnetic moment of m=$3\mu_B$ (S = 3/2). If we reduce the magnetic moment to 1 $\mu_B$(S=1/2) and increase all coupling strengths by 9 times to keep $\alpha S_i^{(z)} \cdot S_j^{(z)}$ as a constant, the calculated $T_c$ does not change at all. This finding confirms the square relation between $T_c$ and magnetic moment. In fact, the similar behavior was also noticed in previous studies about ferroelectricity of monolayer group-IV monochalcogenides. [45] In this sense, our extracted model Eq. (8) can be suitable for general 2D anisotropic Heisenberg systems for both magnetic and electric polarizations, except the S=1/2 systems with only the single-ion anisotropy ($\lambda = 0$). Their spin wave spectrum remains gapless, and magnetic orders may not be maintained, which cannot be captured by classic MC simulations. [28,29]

## VI. Phase diagram of the anisotropic interactions

There is an obvious question about the linear expression of Eq. (8): the different magnetic interactions contribute to $T_c$ independently. This cannot be true for the whole parameter space. The extreme example is that when the anisotropic interactions are zero, that formula still give a finite $T_c$. This obviously conflicts with the MW theorem. [1,46] In other words, this linear expression can only be true within a suitable regime (parameter space). To clarify this point, we study the phase diagram within a much larger parameter space, particularly for the small anisotropic regimes.

To address essential physics and avoid unnecessary complicity of scanning high-dimensional parameter space, we focus on two simplified XXZ Heisenberg models:

$$\mathcal{H}_a = \sum_i A(S_i^z)^2 + \sum_{<i,j>} \frac{1}{2}(J\vec{S}_i \cdot \vec{S}_j) \qquad (9)$$

$$\mathcal{H}_b = \sum_{<i,j>} \frac{1}{2}(\lambda_1 S_i^z S_j^z + J_1 \vec{S}_i \cdot \vec{S}_j) \qquad (10)$$

The first one (Eq. 9) focuses on how the onsite anisotropy impacts magnetic phases, while the second one (Eq. 10) focuses on how the anisotropic coupling impacts magnetic phases. The MC results are presented in Figure 6. Both models exhibit the similar phase diagram. There are basically three phases: the out-of-plane FM phase, the paramagnetic (PM) phase, and the planar phase. The planar phase is from the positive sign of anisotropic terms, which drive magnetization from the out-of-plane easy axis (z) to the in-plane easy plane (x-y). In the following, we mainly focus on the out-of-plane FM and PM phases, in which the sign of anisotropic terms is negative.

The most striking part in the phase diagram of Figure 6 is how the anisotropic term (A or $\lambda$) approaches zero. The MW theorem show that anisotropy is necessary for holding the long-range magnetic order while how the anisotropy quantitatively impacts $T_c$ is not answered. In figure 6, it shows that a minor anisotropy can dramatically, nonlinearly increase $T_c$. Our simulated behaviors around perfect isotropic point agree with previous scaling [47] and renormalization group [48] analysis, in which $T_c$ approaches zero logarithmically as anisotropic term decreases. [49,50] In this sense, anisotropy works as a seed to create the thermodynamically stable magnetic order while other couplings can further enhance the magnetic order. Due to this sharp variation, the region of divergence which adjoins isotropic point is extremely small. Even if we take $\frac{\alpha}{|J_1|} \sim -0.001$ ($\alpha = A, \lambda$), our MC simulation still exhibits a none-zero $T_c$.

On the other hand, this sharp nonlinear region is so narrow that it provides a good chance for estimating $T_c$ by the first-order linear approximation within a reasonable region. More specifically, for most 2D magnetic materials, the amplitudes of anisotropic terms are substantially smaller than that of the isotropic NN coupling ($J_1$), as shown in Table I. Therefore, the true material parameter space is very narrow, which is among the grey-colored space in Figures 6 (a1) and (b1). We amplify that regime and observe a very good linear relation in Figure 6 (a2) and (b2). This is why we obtain the linear relation between $T_c$ and magnetic couplings in Figures 3 and 4. Finally, in Figures 6 (a1) and (b1), when the anisotropy is very large (the far-left side), we approach the MFA limit discussed in section V. A linear relation is also observed but with different slopes from that of realistic materials marked by grey-colored space).

Finally, we must point out that the values of magnetic interactions, i.e., $A$, $J$, and $\lambda$, are crucial for deciding the value of $T_c$, as seen in Eq. 8. Unfortunately, it is known challenging to accurately calculate those magnetic interaction coefficients of correlated materials by *ab initio* approaches. Therefore, how to obtain reliable magnetic interactions is the fundamental challenge for studying these materials and MC simulations, and this is beyond the scope of this work. On the other hand, our proposed linear model survives in a wide range of parameter space (Figures 3 and 4), making it robust as long as reliable magnetic interactions are obtained.

## VII. Conclusions

In this work, we have calculated the Curie temperature $T_c$ of the widely studied monolayer chromium trihalides and CGT by MC simulations based on the XXZ model with magnetic interactions extracted from first-principles calculations. Our calculated $T_c$ of monolayer $CrI_3$ agrees excellently with measurements. Moreover, we find a universal, linear dependence between $T_c$ and magnetic interactions within the parameter space of realistic materials. With this linear dependence, we can predict $T_c$ of 2D magnets without MC simulations once obtaining reliable magnetic interactions. This linear model provides insights to clarify and understand the roles of isotropic and anisotropic magnetic interactions in deciding $T_c$: the anisotropic terms typically ensure the stability for the FM order in 2D magnets, and isotropic terms basically decide the magnitude of $T_c$. It also sheds light on searching for novel 2D materials with higher Curie temperatures by engineering the anisotropic and isotropic magnetic interactions.

**Acknowledgement**

The work is supported by the National Science Foundation (NSF) CAREER Grant No. DMR-1455346 and the Air Force Office of Scientific Research (AFOSR) grant No. FA9550-17-1-0304. The computational resources have been provided by the Stampede of TeraGrid at the Texas Advanced Computing Center (TACC) through XSEDE.

Table I: Extracted magnetic interaction parameters of Eq. (1), $T_c$ of MC simulations, linear-model estimations, and experimental measurements. The unit of parameters is meV.

|  | A | $\lambda_1$ | $J_1$ | $\lambda_2$ | $J_2$ | MC $T_c$ (K) | Model $T_c$ (K) | Exp. $T_c$ (K) Monolayer | Exp. $T_c$ (K) Bulk |
|---|---|---|---|---|---|---|---|---|---|
| $CrI_3$ | -0.087 | -0.085 | -2.12 | 0.02 | -0.35 | 42.2 | 42.8 | 45K [6] | 61K [39] |
| $CrBr_3$ | -0.02 | -0.016 | -1.35 | -0.001 | -0.153 | 23.1 | 24.0 | 27K [42] | 37K [51] |
| $CrCl_3$ | -0.007 | -0.002 | -0.79 | 0 | -0.071 | 12.1 | 13.1 | 17K(2L) [42,43] | 17K [40] |
| $Cr_2Ge_2Te_6$* | -0.01 | - | -2.71 | - | 0.0058 | - | 34.5 | 30K(2L) [7] | 68K [7] |

Table II: Extracted coefficients of the linear model in Eq (8) based on MC simulations.

|  | $\alpha_0$ (A) | $\alpha_1^z$ ($\lambda_1$) | $\alpha_1$ ($J_1$) | $\alpha_2^z$ ($\lambda_2$) | $\alpha_2$ ($J_2$) | $n_1$ | $n_2$ |
|---|---|---|---|---|---|---|---|
| Hexagonal | -0.33 | -0.28 | -0.164 | -0.35 | -0.256 | 3 | 6 |
| Square | -0.34 | -0.29 | -0.183 | -0.35 | -0.257 | 4 | 4 |

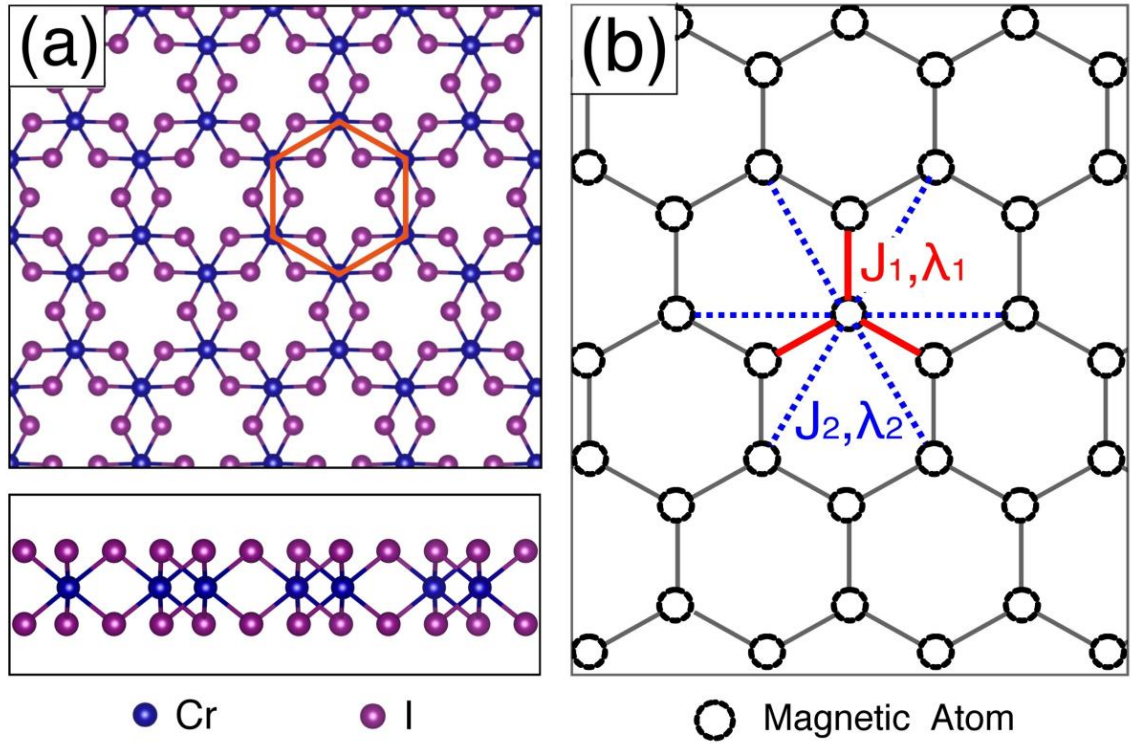

Figure 1. (a) The top and side views of the atomic structure of monolayer $CrI_3$. The magnetic Cr atoms form hexagonal lattices. (b) The XXZ model applies to hexagonal lattices. Each site has a localized magnetic moment, and the NN and NNN coupling interactions are indicated.

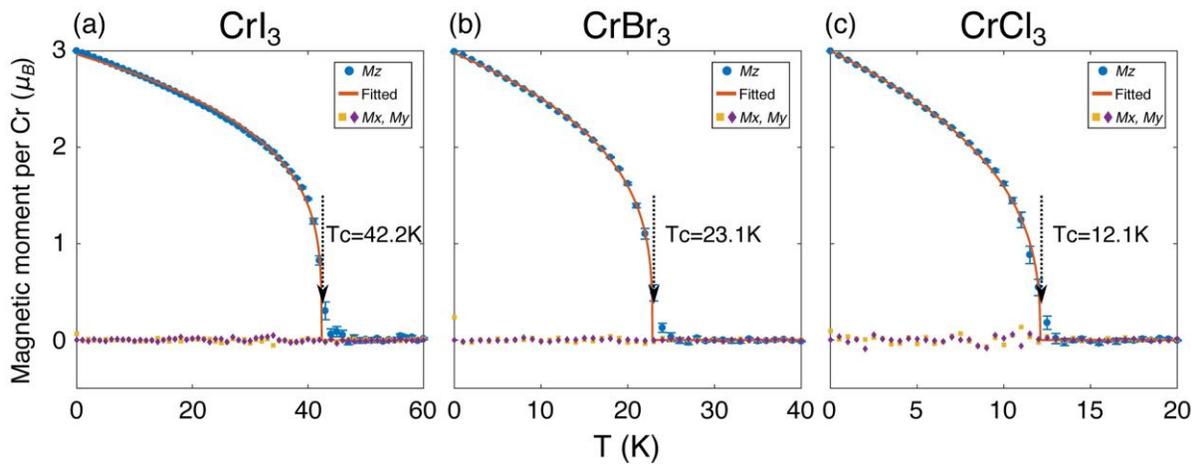

Figure 2 The MC simulated magnetism versus temperature for monolayer $CrX_3$ (X=I, Br, Cl) with the error bar.

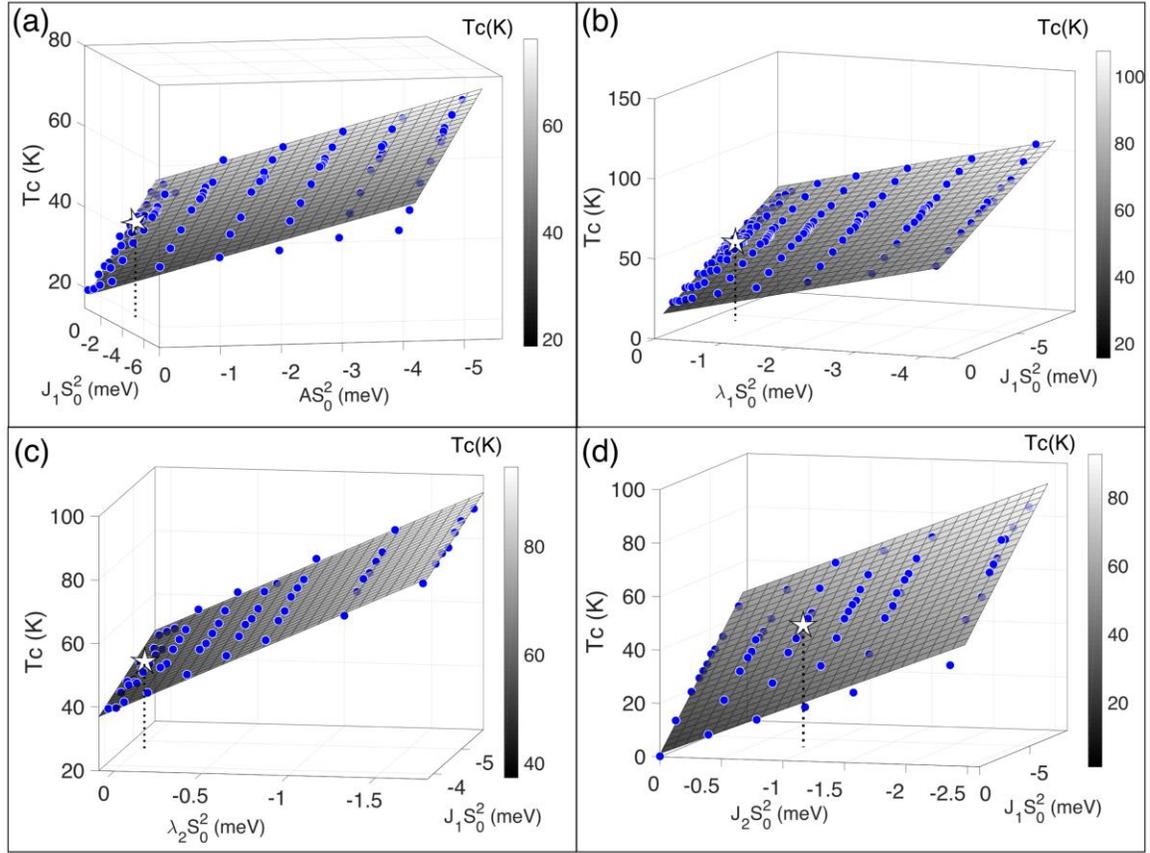

Figure 3. The fitted linear relation between $T_c$ and corresponding parameters of the XXZ model of 2D hexagonal lattices. The MC simulation points are sold blue points. (a-d) are the linearly fitted planes for corresponding onsite anisotropic A and NN isotropic $J_1$, NN anisotropic coupling $\lambda_1$ and $J_1$, NNN anisotropic coupling $\lambda_2$ and $J_1$, and NNN isotropic coupling $J_2$ and $J_1$, respectively. The corresponding parameters of monolayer CrI$_3$ are marked as white stars.

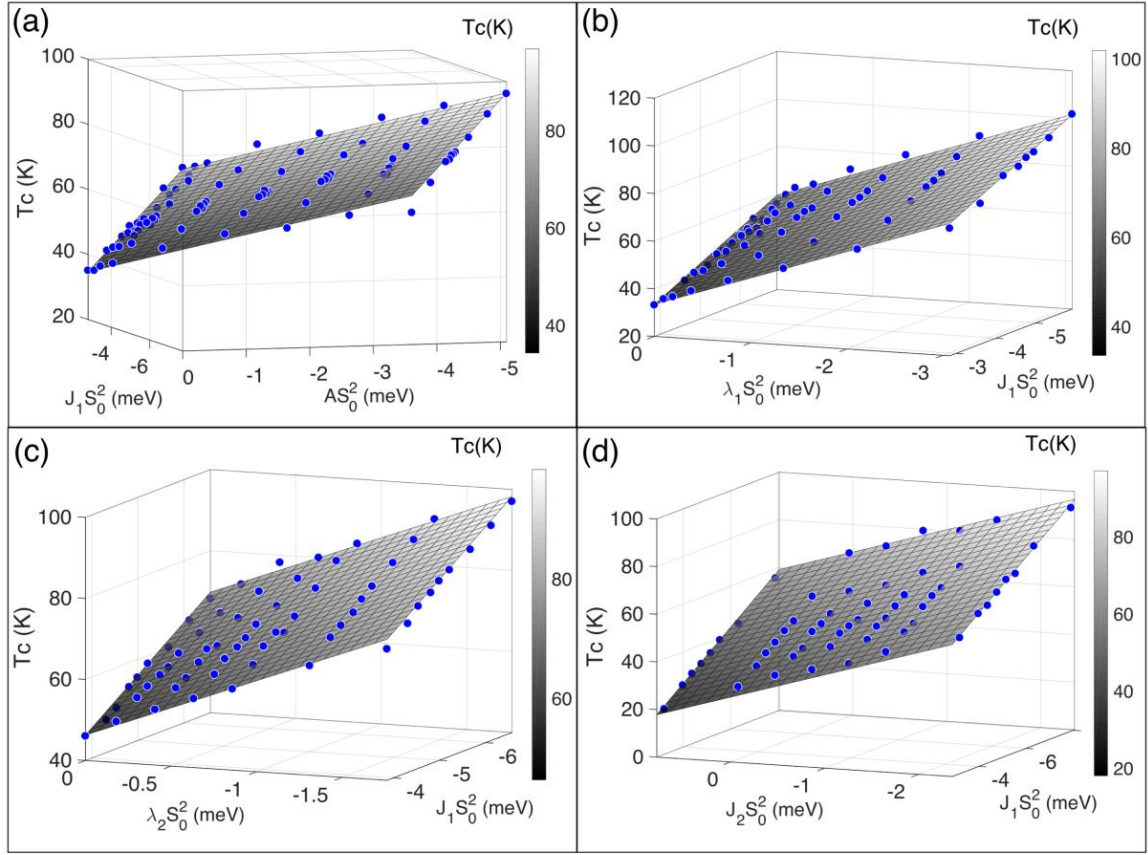

Figure 4. The fitted linear relation between $T_c$ and corresponding parameters of the XXZ model of 2D square lattices. The MC simulation points are sold blue points. (a-d) are the linearly fitted planes for corresponding onsite anisotropic A and NN isotropic $J_1$, NN anisotropic coupling $\lambda_1$ and $J_1$, NNN anisotropic coupling $\lambda_2$ and $J_1$, and NNN isotropic coupling $J_2$ and $J_1$, respectively.

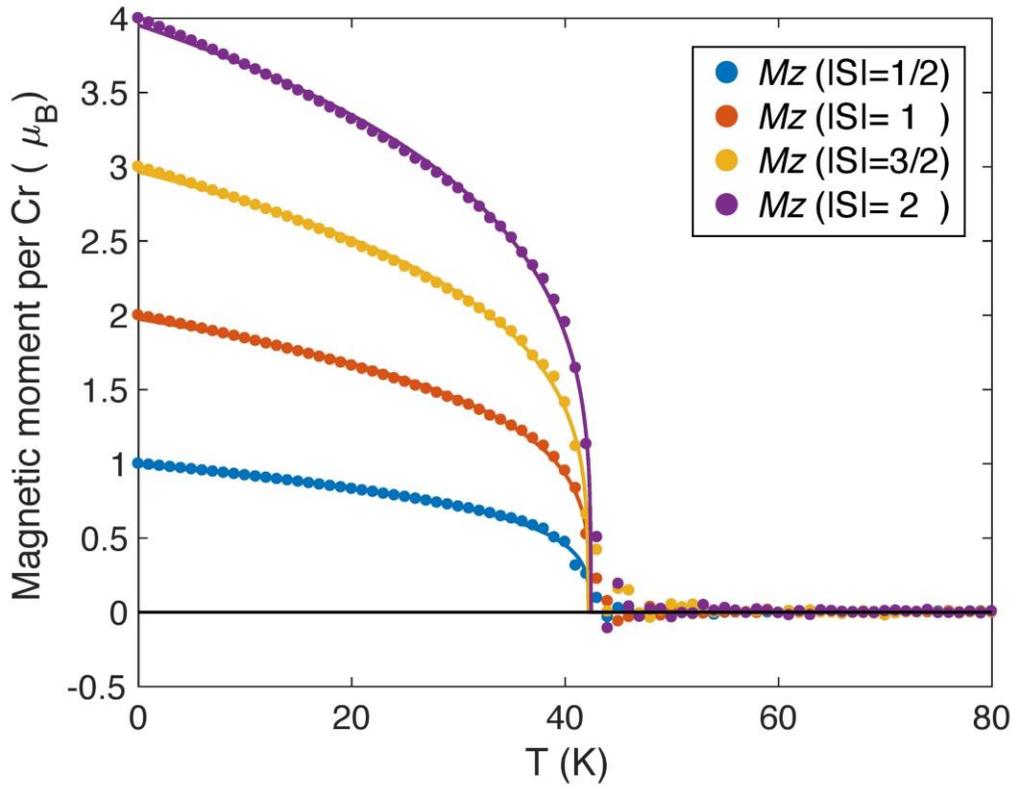

Figure 5 The MC simulated magnetism versus temperature for different magnetic-moment systems. The products $\alpha S_i^{(z)} \cdot S_j^{(z)}$ ($\alpha = A$, $\lambda$, $J$) are fixed at the same value for different magnetic moments.

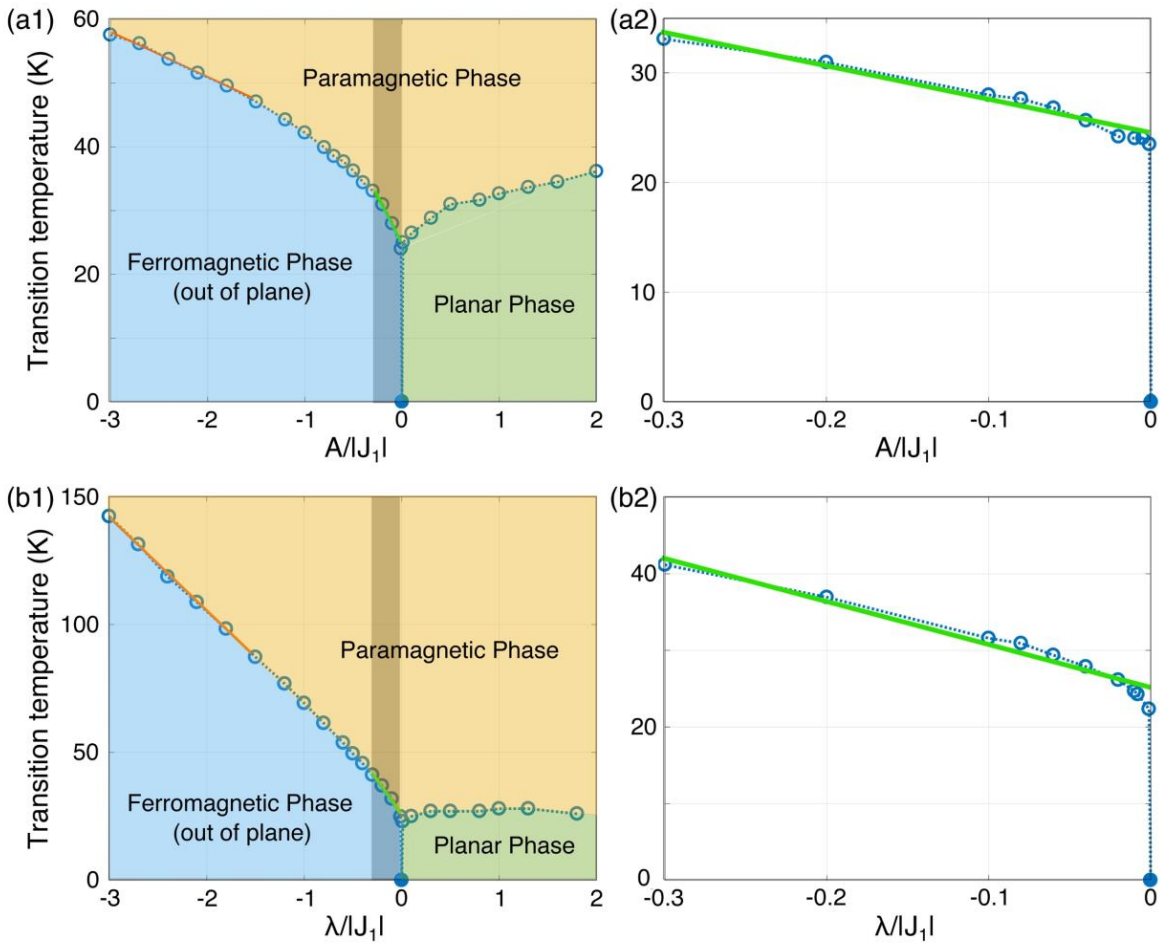

Figure 6 (a1) and (b1) are phase diagrams generated from MC simulations of the Hamiltonian of Eq. (9) and Eq. (10), respectively. The open circles are MC simulated phase boundaries. (a2) and (b2) are amplified from the shaded area with more data points.